\documentclass[dvipsnames]{article} %
\usepackage{colm2024_conference}
\pdfoutput=1
\usepackage{natbib}
\setcitestyle{authoryear,round,citesep={;},aysep={,},yysep={;}}
\let\cite\citep
\usepackage{booktabs}
\usepackage{graphicx}
\usepackage{enumitem}
\usepackage{wrapfig}
\usepackage{algorithm}
\usepackage{algpseudocode}
\usepackage{microtype}
\usepackage{amsmath}
\usepackage{amsthm}
\usepackage{colortbl}
\usepackage[utf8]{inputenc}
\definecolor{lightgray}{rgb}{0.9,0.9,0.9}
\usepackage{caption}
\usepackage{subcaption}
\usepackage{setspace}
\usepackage{url}
\usepackage{multirow}
\usepackage{colortbl}
\usepackage{tabularx}
\usepackage{blindtext}
\usepackage{pgfplots}
\pgfplotsset{compat=1.18} 
\usepackage{tikz}
\usetikzlibrary{er,positioning,bayesnet}
\usepackage{makecell}
\usepackage{tipa}
\usepackage{siunitx}
\usepackage{nicefrac}
\usepackage{tocloft}
\usepackage{listings}
\usepackage{cleveref}
\usepackage{placeins}
\usepackage{rotating}
\usepackage{silence}
\usepackage{tablefootnote}
\usepackage{threeparttable}
\usepackage[raster,skins]{tcolorbox} %
\usepackage{xltabular}
\usepackage{adjustbox}
\usepackage{xurl}
\usepackage[normalem]{ulem}
\useunder{\uline}{\ul}{}
\usepackage{newtxtext,newtxmath}
\usepackage{CJKutf8}
\usepackage{pifont}
\usepackage{xcolor}

\definecolor{darkgreen}{rgb}{0.0, 0.6, 0.0}
\newcommand{\up}[1]{\rlap{\textsuperscript{\textcolor{red}{\scriptsize$\uparrow$#1}}}}
\newcommand{\down}[1]{\rlap{\textsuperscript{\textcolor{darkgreen}{\scriptsize$\downarrow$#1}}}}
\definecolor{pyellow}{rgb}{1.0, 1.0, 0.8} 

\usepackage[table]{xcolor}
\definecolor{pyellow}{RGB}{255,245,200}

\usepackage{amsmath,amsfonts,bm}

\def\eqref#1{equation~\ref{#1}}

\def\1{\bm{1}}

\DeclareMathAlphabet{\mathsfit}{\encodingdefault}{\sfdefault}{m}{sl}
\SetMathAlphabet{\mathsfit}{bold}{\encodingdefault}{\sfdefault}{bx}{n}

\newcommand*\justify{%
  \fontdimen2\font=0.4em%
  \fontdimen3\font=0.2em%
  \fontdimen4\font=0.1em%
  \fontdimen7\font=0.1em%
  \hyphenchar\font=`\-%
}

\renewcommand{\texttt}[1]{%
  \begingroup
  \ttfamily
  \begingroup\lccode`~=`/\lowercase{\endgroup\def~}{/\discretionary{}{}{}}%
  \begingroup\lccode`~=`[\lowercase{\endgroup\def~}{[\discretionary{}{}{}}%
  \begingroup\lccode`~=`.\lowercase{\endgroup\def~}{.\discretionary{}{}{}}%
  \catcode`/=\active\catcode`[=\active\catcode`.=\active
  \justify\scantokens{#1\noexpand}%
  \endgroup
}

\title{CORD: Bridging the Audio–Text Reasoning Gap via Weighted On-policy Cross-modal Distillation}

\author{
 \centering
 \small{}
\textbf{Jing Hu$^{*1,2}$  \hspace{4mm} Danxiang Zhu$^{*1}$  \hspace{4mm} 
Xianlong Luo$^{1}$  \hspace{4mm} Dan Zhang$^{1}$ \hspace{4mm} Shuwei He$^{1,3}$ \hspace{4mm} Yishu Lei$^{1}$ \hspace{4mm} \\ Haitao Zheng$^{2}$ \hspace{4mm}  Shikun Feng$^{\dag1}$ \hspace{4mm} Jingzhou He$^{1}$ \hspace{4mm} Yu Sun$^{1}$ \hspace{4mm} Hua Wu$^{1}$ \hspace{4mm} Haifeng Wang$^{1}$} \\
\vspace{1em}
 \centering
 \small{}
  $^{1}$ERNIE Team, Baidu \\
  $^{2}$Tsinghua Shenzhen International Graduate School, Tsinghua University \\
  $^{3}$College of Computer Science, Inner Mongolia University \\
  \vspace{1em}
   \centering
   \small{}
  \texttt{{cminuser@gmail.com} {zhudanxiang@baidu.com} \\\{luoxianlong, zhangdan20, heshuwei, leiyishu\}@baidu.com \\ 
  zheng.haitao@sz.tsinghua.edu.cn \\
  \{fengshikun01, hejingzhou, sunyu02, wu\_hua, wanghaifeng\}@baidu.com}
  \\
  \vspace{1em}
  $^{*}$Equal contribution, $^{\dagger}$Corresponding author
}

\makeatletter\def\@abstract{
Large Audio Language Models~(LALMs) have garnered significant research interest.
Despite being built upon text-based large language models~(LLMs), LALMs frequently exhibit a degradation in knowledge and reasoning capabilities.
We hypothesize that this limitation stems from the failure of current training paradigms to effectively bridge the acoustic-semantic gap within the feature representation space.
To address this challenge, we propose CORD, a unified alignment framework that performs online cross-modal self-distillation.
Specifically, it aligns audio-conditioned reasoning with its text-conditioned counterpart within a unified model.
Leveraging the text modality as an internal teacher, CORD performs multi-granularity alignment throughout the audio rollout process.
At the token level, it employs on-policy reverse KL divergence with importance-aware weighting to prioritize early and semantically critical tokens.
At the sequence level, CORD introduces a judge-based global reward to optimize complete reasoning trajectories via Group Relative Policy Optimization~(GRPO).
Empirical results across multiple benchmarks demonstrate that CORD consistently enhances audio-conditioned reasoning and substantially bridges the audio–text performance gap with only 80k synthetic training samples, validating the efficacy and data efficiency of our on-policy, multi-level cross-modal alignment approach.
}\makeatother

\begin{document}
\maketitle

\pagestyle{firstpage}  %
\pagestyle{normalpage}

\section{Introduction}

Large Language Models (LLMs)~\citep{qwen2025qwen25technicalreport,deepseekai2025deepseekv3technicalreport,openai2025gptoss120bgptoss20bmodel} have demonstrated exceptional semantic understanding capabilities, sparking research to extend this intelligence to multimodal domains through end-to-end processing. In the audio domain, most state-of-the-art Large Audio-Language Models (LALMs) \cite{chu2023qwenaudioadvancinguniversalaudio,chu2024qwen2,wu2025step,ding2025kimi,zeng2024glm,fang2024llamaomniseamlessspeechinteraction} are built upon pretrained text-based LLMs by incorporating an audio encoder~\citep{radford2022robustspeechrecognitionlargescale} and a modality alignment module. Raw audio signals are first encoded into acoustic representations, which are then projected into the LLM’s embedding space through audio-text paired supervision. This paradigm implicitly assumes that training on audio-text interleaved data is sufficient to align audio and text into a unified semantic space.

However, recent studies~\citep{wang2024blspbootstrappinglanguagespeechpretraining, cuervo2025closing} have revealed a persistent performance disparity between the two modalities. Despite receiving semantically equivalent inputs, LALMs often exhibit markedly inferior performance on audio-conditioned tasks compared to their text-conditioned counterparts. This gap is particularly pronounced in data-constrained regimes~\citep{chu2024qwen2,wu2025step,xu2025qwen25omnitechnicalreport}, where limited training samples expose the limitations of existing alignment mechanisms and highlight the need for more data-efficient cross-modal alignment.

Several recent approaches have attempted to improve audio-conditioned reasoning performance, primarily through supervised fine-tuning with labeled speech data or knowledge distillation from external text-based teachers. Despite their progress, these methods suffer from three fundamental limitations. 
\textbf{(1) Limited scalability.} Supervised fine-tuning \citep{he2024emiliaextensivemultilingualdiverse,minixhofer2025scaling} relies on large-scale, high-quality annotated speech data, which is expensive to collect and difficult to scale across diverse tasks and domains. 
\textbf{(2) Off-policy distillation and distribution mismatch.} Teacher-based distillation methods \cite{cuervo2025closing,wang2024blspbootstrappinglanguagespeechpretraining,tseng2025tastetextalignedspeechtokenization} typically provide supervision along the teacher’s text-generation trajectories, rather than the student’s actual audio-conditioned inference states. This off-policy supervision leads to a distribution mismatch and limits the ability to correct accumulated audio-specific reasoning errors. 
\textbf{(3) Uniform token-level supervision.} Conventional KL-based distillation treats all tokens equally~\citep{wang20258020rulehighentropyminority,li2025attentionilluminatesllmreasoning,tseng2025tastetextalignedspeechtokenization}, failing to emphasize semantically critical tokens that drive cross-modal misalignment, and lacking sequence-level constraints to explicitly regulate global reasoning trajectories.

To address these challenges, we propose \textbf{CORD} (\textbf{C}ross-modal Weighted \textbf{O}n-policy \textbf{R}eward-guided \textbf{D}istillation), a unified alignment framework that performs online cross-modal self-distillation without relying on any external teacher. CORD leverages the model’s internal text modality as an in-model teacher and conducts multi-granularity distillation along the student model’s real audio-conditioned reasoning trajectories. By constructing on-policy alignment objectives based on audio modality rollouts, CORD directly aligns audio-conditioned reasoning behavior with its text-conditioned counterpart within a unified model architecture.

At the token level, CORD introduces a fine-grained weighting mechanism that prioritizes tokens exhibiting high cross-modal divergence as well as those appearing at early stages of reasoning, where semantic deviations are more likely to propagate and dominate the final outcome. This weighted reverse KL objective enables targeted correction of modality-specific reasoning errors. At the sequence level, CORD formulates a cross-modal reward function and employs Group Relative Policy Optimization (GRPO) \citep{shao2024deepseekmath} to align complete reasoning trajectories, encouraging the audio modality to follow reasoning policies consistent with the text modality.

Extensive experiments on multiple reasoning benchmarks demonstrate that CORD substantially improves audio-conditioned reasoning performance. In particular, CORD reduces the audio--text performance gap by an average of \textbf{41.6\%} on Qwen2-Audio-7B-Instruct and \textbf{44.8\%} on Step-Audio2-mini, significantly outperforming conventional distillation baselines. On several tasks, CORD nearly eliminates the modality gap altogether, indicating that on-policy cross-modal self-distillation with an internal teacher provides an effective and scalable solution for mitigating reasoning degradation in LALMs.

Our contributions can be summarized as follows:
\begin{itemize}
    \item \textbf{Internal on-policy cross-modal self-distillation.} 
    We propose \textbf{CORD}, a fully in-model, weighted on-policy cross-modal self-distillation framework that aligns audio-conditioned reasoning with text-conditioned behavior without relying on any external teacher, avoiding off-policy mismatch and architecture-induced noise.

    \item \textbf{Multi-granularity alignment of reasoning trajectories.} 
    CORD jointly enforces token-level and sequence-level alignment by emphasizing semantically critical and early reasoning tokens with weighted reverse KL, while regulating global reasoning trajectories via a reward-guided GRPO objective.

    \item \textbf{Effective reduction of the audio-text reasoning gap.} 
    We demonstrate that CORD consistently reduces the audio-text gap by over 40\% across various backbones, nearly reaching parity with text-conditioned performance in several reasoning tasks.
\end{itemize}

\begin{figure*}
    \centering
    \includegraphics[width=1\linewidth]{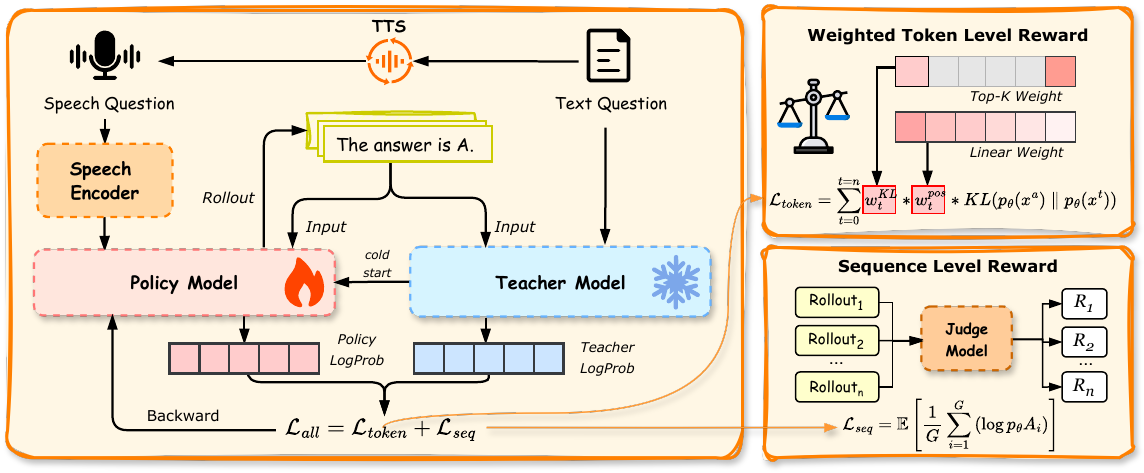}
    \caption{The overall framework of \textbf{CORD}. Given semantically equivalent audio and text inputs, CORD performs on-policy cross-modal self-distillation within a single model. Audio-conditioned trajectories are aligned to text-conditioned behaviors at two levels: (i) a token-level objective that applies importance-aware and position-aware reverse KL weighting along audio rollouts, and (ii) a sequence-level objective that uses a judge-based reward optimized via GRPO to enforce global reasoning consistency.}
    \label{fig:placeholder}
\end{figure*}

\section{Related Work}

\subsection{Audio-Text Alignment in LALMs}
While cascaded ASR–LLM pipelines largely preserve text-domain performance, they discard speaker and paralinguistic cues essential for speech interaction \citep{maimon25salmon}
, prompting recent work on end-to-end LALMs \citep{tang2024salmonn,chu2024qwen2, xie2024miniomnilanguagemodelshear}.
However, despite architectural advances, such models consistently underperform text-based LLMs on language understanding and reasoning benchmarks, revealing a persistent \emph{text–speech understanding gap}~\citep{cui-etal-2025-voxeval}
. Existing approaches mainly attempt to reduce this gap via representation-level cross-modal alignment or large-scale synthetic speech data augmentation~\citep{held2024distillingendtoendvoiceassistant}, but often show limited gains on broad reasoning tasks or rely on massive proprietary datasets that hinder reproducibility ~\citep{zeng2024scalingspeechtextpretrainingsynthetic}

\subsection{Knowledge Distillation}
Standard knowledge distillation (KD)~\cite{10.1145/1150402.1150464,hinton2015distillingknowledgeneuralnetwork} for Large Language Models typically employs a teacher model to generate training trajectories. The student model is then supervised to match the teacher's output distribution, either by minimizing the forward KL divergence on the teacher's logits or by using cross-entropy loss to learn from the sampled tokens. However, recent works~\citep{agarwal2024onpolicydistillationlanguagemodels,gu2025minillmknowledgedistillationlarge} have introduced On-Policy Distillation (OPD). This paradigm shifts the focus from teacher-generated paths to trajectories sampled directly from the student's current policy, enabling the model to learn from states it actually encounters during inference.
Building on this, recent studies~\citep{yang2025qwen3technicalreport,lu2025onpolicydistillation,chen2025retainingdoingroleonpolicy} demonstrate that OPD can achieve reasoning performance comparable to RL-style alignment with significantly lower computational overhead. Furthermore, this paradigm has been extended to black-box settings~\citep{ye2025blackboxonpolicydistillationlarge}.

\section{Method}
\label{sec:method}

In this section, we present CORD, a framework designed to bridge the performance gap in LALMs by aligning audio-conditioned reasoning trajectories with their text-conditioned counterparts. Unlike static or off-policy supervision, CORD enforces alignment along the model's actual inference paths using an in-model teacher strategy. As shown in Figure~\ref{fig:placeholder}, CORD employs a multi-granularity alignment approach consisting of two complementary objectives:

\begin{itemize}
\item \textbf{Token-level alignment} corrects fine-grained semantic deviations at each decoding step using importance-weighted KL.
\item \textbf{Sequence-level alignment} enforces global consistency across entire trajectories via GRPO.
\end{itemize}

By combining local step-wise correction with global trajectory regulation, CORD provides a unified framework for cross-modal alignment. In the following, we detail the problem formulation, the specific alignment objectives, and provide a comprehensive analysis of the underlying mechanisms that ensure training stability and semantic fidelity.

\subsection{Problem Formulation}
Given semantically equivalent audio and text inputs $(x^{a}, x^{t})$, a LALM $p_{\theta}$ induces autoregressive distributions over output sequences $y$:
\begin{equation}
p_{\theta}(y\mid x) = \prod_{t=1}^{T} p_{\theta}(y_t \mid y_{<t}, x), \quad x \in \{x^{a}, x^{t}\}.
\end{equation}
Empirically, LALMs often exhibit divergent behaviors across modalities despite identical input semantics. CORD aims to minimize this discrepancy by explicitly constraining the \emph{audio-conditioned inference trajectories} $y \sim p_{\theta}(\cdot \mid x^{a})$, rather than merely matching marginal output distributions.

\begin{figure}
    \centering
    \begin{subfigure}[t]{1\linewidth}
    \includegraphics[width=\linewidth]{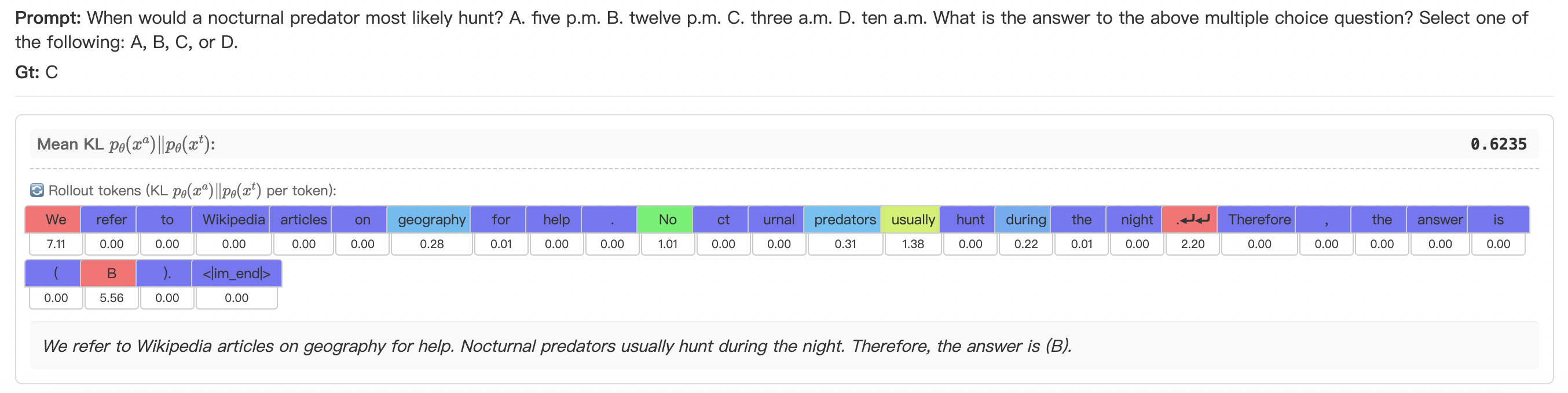}
    \caption{Incorrect case: the model selects an incorrect option.}
    \label{fig:kl_vis_a}
    \end{subfigure}
    \hfill
    \begin{subfigure}[t]{1\linewidth}
    \includegraphics[width=\linewidth]{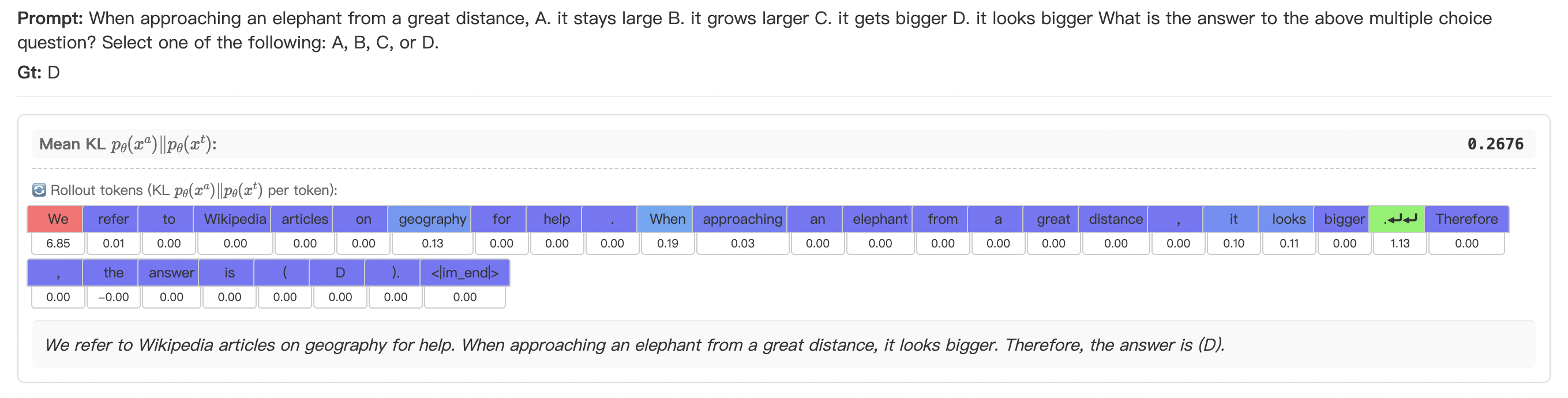}
    \caption{Correct case: the model selects the correct option.}
    \label{fig:kl_vis_b}
    \end{subfigure}
    
\caption{
Token-level reverse KL visualization along audio-conditioned rollouts generated by \emph{Qwen2-Audio-7B-Instruct}.
Each block represents a generated token, annotated with its reverse KL divergence between text- and audio-conditioned distributions.
The correct case shows low divergence on the final option token, while the incorrect case exhibits large reverse KL at the decision token and several early reasoning steps.
Notably, the incorrect output contains formulaic phrases such as ``\emph{We refer to $\cdots$},'' reflecting a hallucination-prone pattern under audio inputs.
}
    \label{fig:kl_visualization}
\end{figure}

\subsection{On-policy Cross-modal Distillation}
As shown in Figure~\ref{fig:placeholder}, CORD aligns the model's cross-modal behavior along trajectories sampled from its current policy. For each $x^{a}$, we sample $y \sim p_{\theta}(\cdot \mid x^{a})$ to obtain an on-policy decoding trajectory. At each step $t$, the model induces two distributions over the vocabulary $\mathcal{V}$ conditioned on the same prefix $y_{<t}$: $p_{\theta}(\cdot \mid y_{<t}, x^{a})$ and $p_{\theta}(\cdot \mid y_{<t}, x^{t})$. 

To quantify the discrepancy at each state, we employ \textbf{Reverse KL Divergence}:
\begin{equation}
\begin{aligned}
D_t
&=
\mathrm{KL}\left( 
p_{\theta}(\cdot \mid y_{<t}, x^a)
\ \| \ 
p_{\theta}(\cdot \mid y_{<t}, x^{t})
\right) 
\\
&=
\sum_{v \in \mathcal{V}}
p_{\theta}(v \mid y_{<t}, x^{a})
\log
\frac{
p_{\theta}(v \mid y_{<t}, x^{a})
}{
p_{\theta}(v \mid y_{<t}, x^{t})
}.
\end{aligned}
\label{eq:reverse_kl}
\end{equation}

Compared to forward KL, reverse KL places stronger emphasis on high-probability tokens under the text-conditioned distribution, encouraging the audio-conditioned policy to recover critical reasoning decisions made by the text modality.
When applied \emph{on-policy} along audio-conditioned trajectories, this formulation enables targeted correction of semantic deviations that arise during audio inference, rather than enforcing global distributional matching.

\begin{figure*}[t]
\centering

\begin{minipage}[t]{0.48\linewidth}
\vspace{0pt}
\centering
\includegraphics[width=\linewidth]{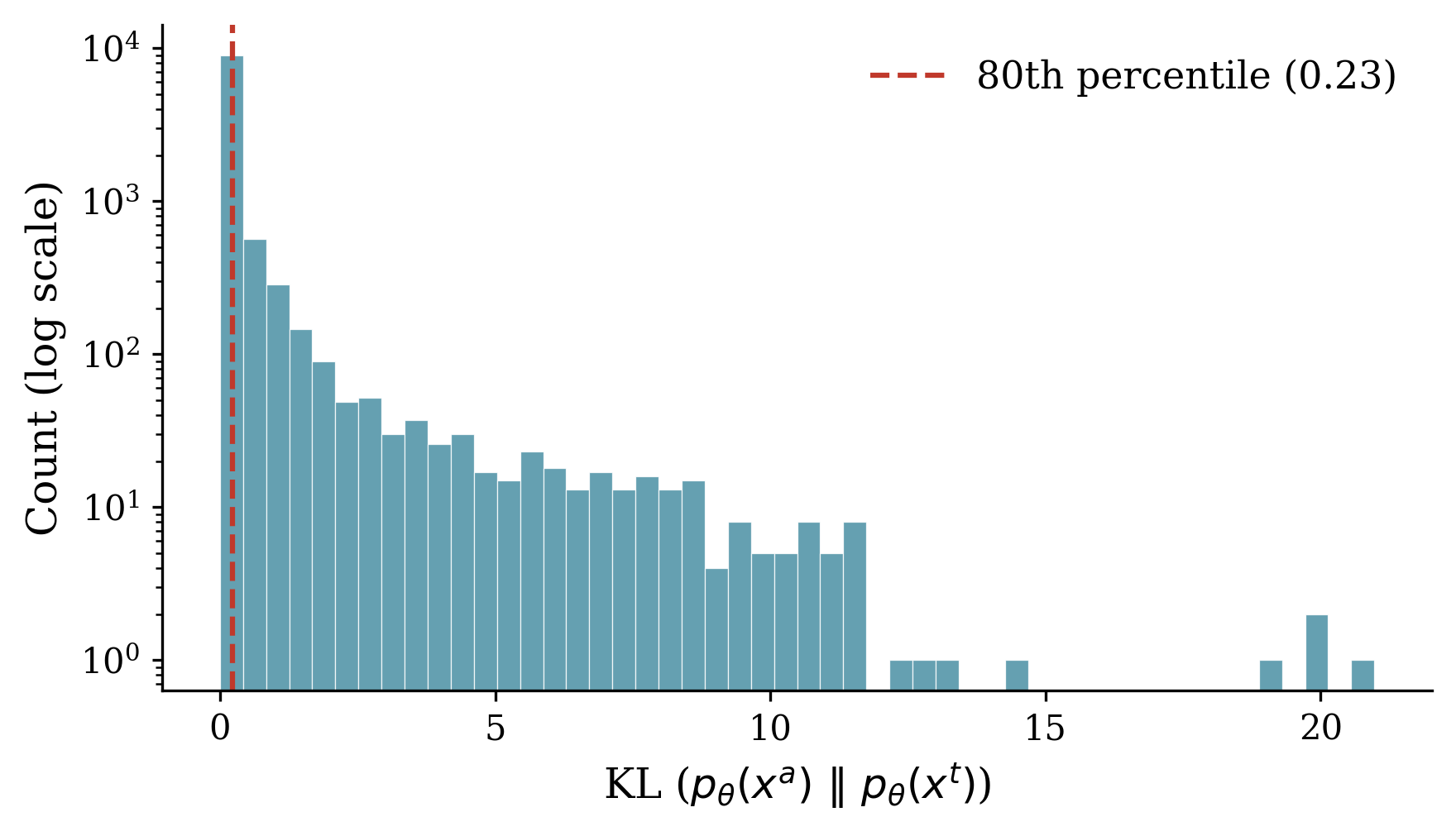}
\includegraphics[width=\linewidth]{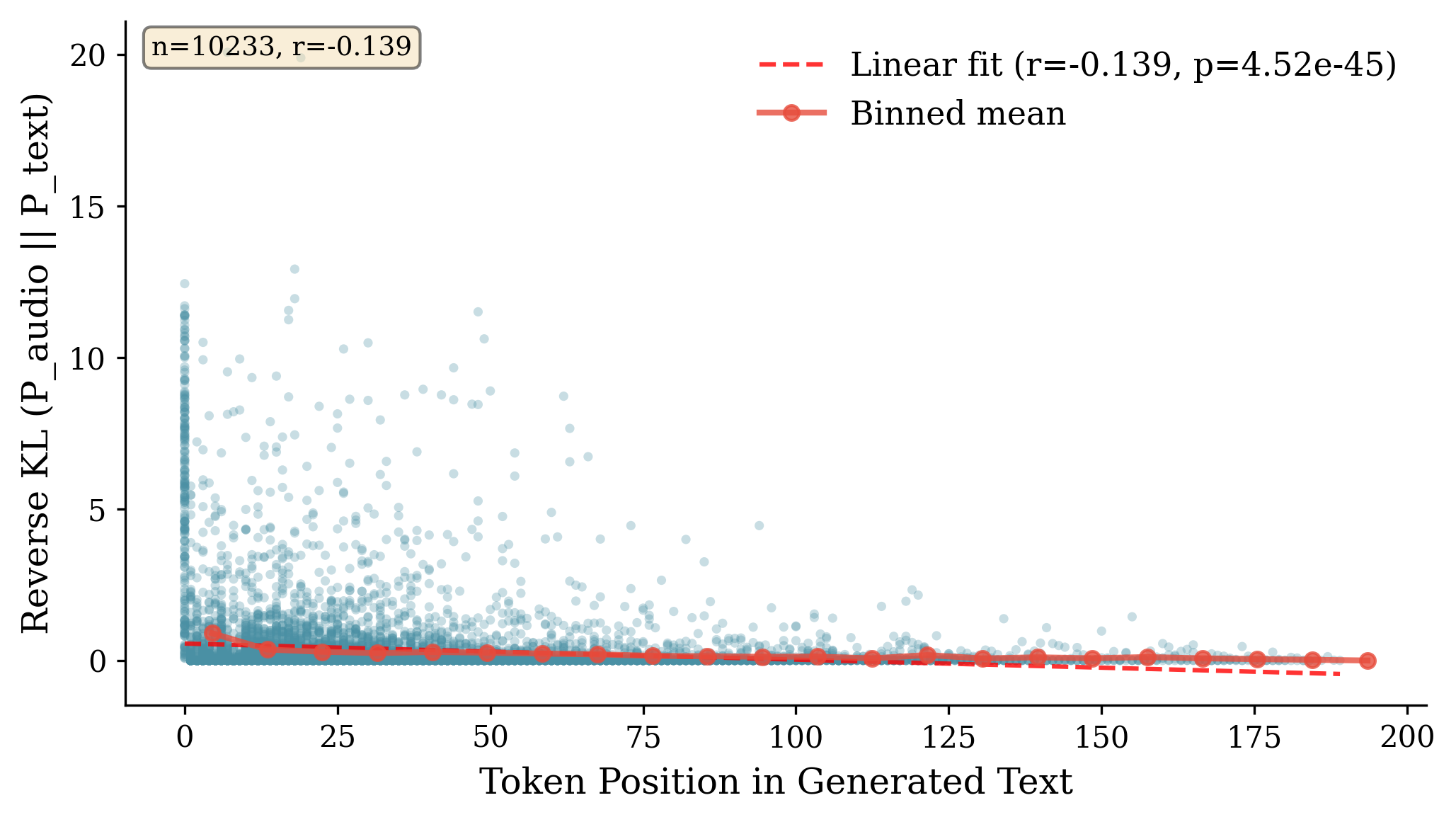}
\vspace{-2pt}
\end{minipage}
\hfill
\begin{minipage}[t]{0.48\linewidth}
\vspace{1em}
\centering
\includegraphics[width=\linewidth]{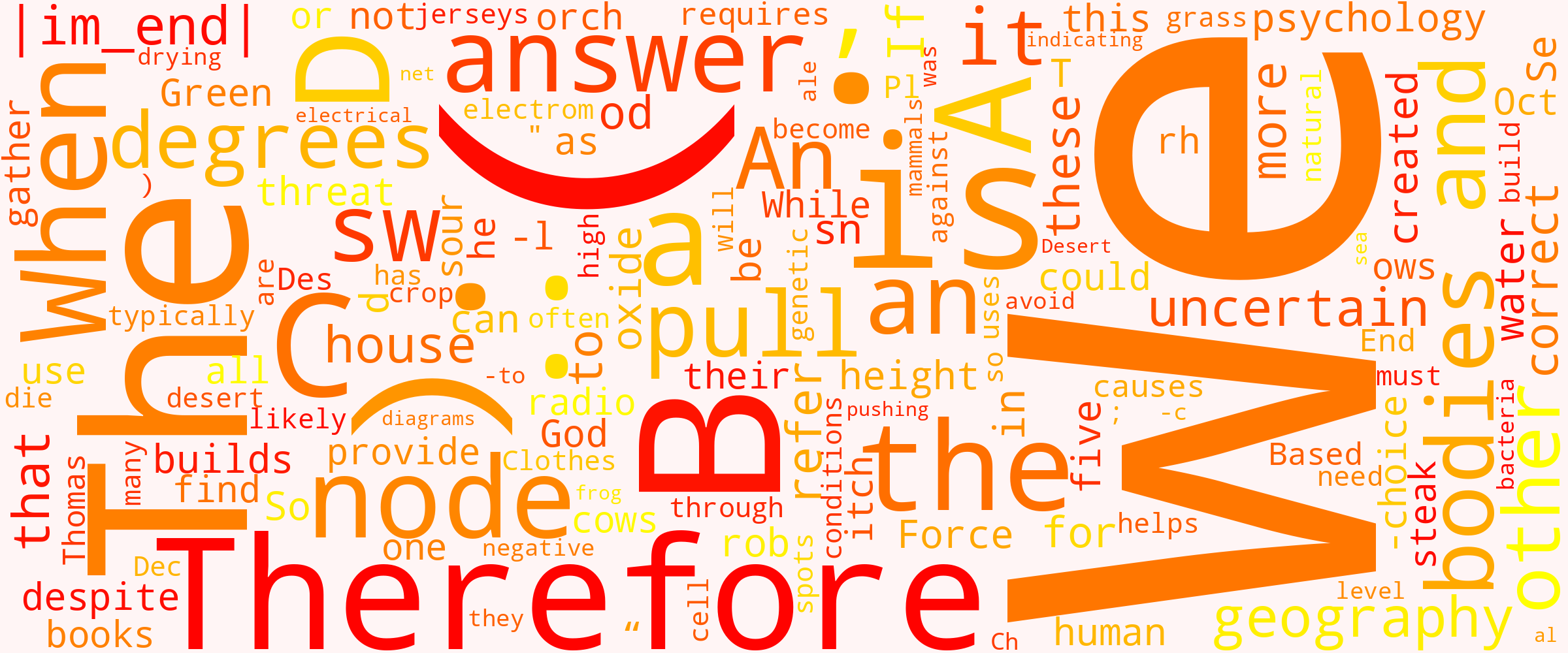}
\subcaption{Tokens from high-KL regions.}
\vspace{2em}

\includegraphics[width=\linewidth]{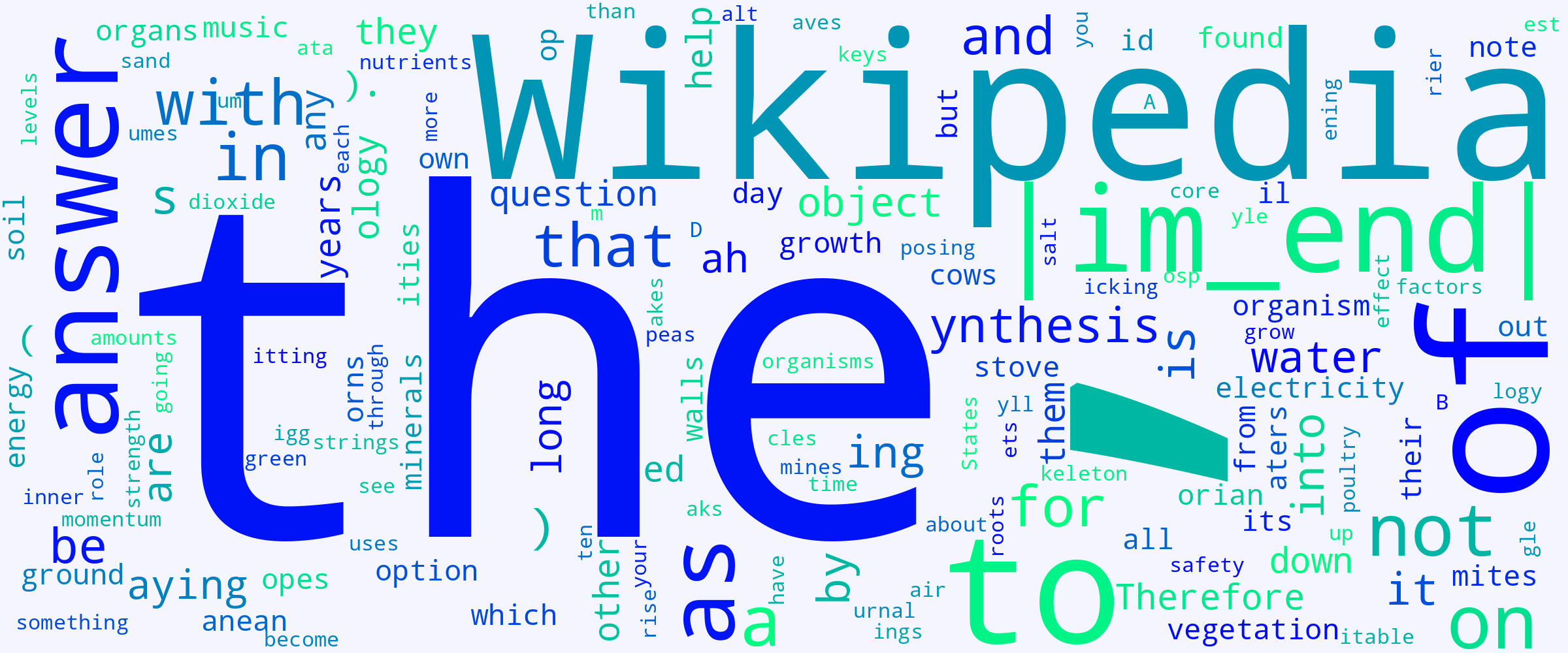}
\subcaption{Tokens from low-KL (bottom) regions.}
\label{fig:kl_compare}
\end{minipage}
\vspace{-1em}
\caption{Statistical and Semantic Analysis of Token-Level Reverse KL Divergence.
Top-left: Histogram of $\text{KL}(p_{\theta}(x^a) \| p_{\theta}(x^t))$ for all generated tokens (log scale), with the 80th percentile threshold marked by the red dashed line. Higher KL values highlight pivotal states where audio-conditioned reasoning significantly deviates from the text-conditioned teacher. Bottom-left: Scatter plot of KL divergence versus token position, revealing that major cross-modal discrepancies are concentrated in earlier reasoning stages. (a)-(b): Qualitative word cloud visualizations. High-KL regions (a) are enriched with semantically critical reasoning tokens (e.g., “Therefore”, “answer”) and choice options (e.g., A, B), whereas low-KL regions (b) primarily consist of common functional words and background information.}
\label{fig:wordcloud}
\end{figure*}

\subsection{Analysis of Cross-modal Discrepancy}
\label{subsec:discrepancy_analysis}
To motivate the design of CORD, we first investigate the reasoning behavior of LALMs under audio and text modalities.
Let $y = (y_1,\dots,y_T)$ denote a sequence sampled from the audio-conditioned policy
$p_\theta(\cdot \mid x^a)$.
At each decoding step $t$, given the same prefix $y_{<t}$, the model induces two
conditional token distributions over the vocabulary $\mathcal{V}$:
one conditioned on audio input $x^a$, and the other on text input $x^t$.
We quantify their discrepancy using the token-level KL divergence $D_t$.

A common practice is to uniformly average token-level KL divergences,
\begin{equation}
\frac{1}{T}\sum_{t=1}^T D_t.
\label{eq:uniform_kl}
\end{equation}
This approach implicitly assumes that all tokens contribute equally to cross-modal alignment. However, our empirical analysis on the MMSU benchmark reveals that the distribution of $D_t$ is highly skewed and non-uniform.

As illustrated in Figure~\ref{fig:wordcloud}, the distribution of $D_t$ is highly skewed: most tokens exhibit very small divergence, while only a small number of critical tokens have substantially larger KL values. 
Specifically, we visualize the reverse KL distribution on the MMSU benchmark to investigate this phenomenon. As illustrated in Figure~\ref{fig:wordcloud} (top-left), the distribution follows a heavy-tailed pattern, where the 80th percentile corresponds to a remarkably low divergence of only 0.23. 
This indicates that the vast majority of tokens are already well-aligned across modalities. 
Furthermore, the scatter plot (bottom-left) shows that these high-discrepancy states are concentrated in early decoding stages ($r = -0.139$). Misalignments at these pivotal early states often trigger a cascade effect, leading to cumulative reasoning failures.

In contrast, the word clouds in Figure~\ref{fig:wordcloud}(a) reveal that high-KL tokens are predominantly concentrated on semantically critical reasoning words and multiple-choice options (e.g., A, B). The misalignment at these pivotal states is a primary driver of incorrect responses; moreover, errors occurring at these early high-discrepancy tokens tend to trigger a cascade effect, leading to cumulative failures in subsequent decoding steps.

To further illustrate this phenomenon at the instance level,
Figure~\ref{fig:kl_visualization} visualizes the token-level reverse KL divergence along audio-conditioned rollouts for two examples.
Figure~\ref{fig:kl_vis_a} results in a correct prediction, while Figure~\ref{fig:kl_vis_b} leads to an incorrect answer.
We observe that the reverse KL on the final option token differs dramatically between the two cases.
More importantly, the incorrect case already exhibits noticeably larger divergence at several early reasoning tokens,
indicating that cross-modal misalignment emerges early and propagates to the final decision.

As a result, uniform averaging leads to a low overall loss magnitude,
causing gradients from high-KL tokens to be diluted by numerous low-KL tokens.
This weakens corrective updates precisely at positions that dominate semantic misalignment.
To address this issue, we introduce an importance-aware token-level weighting scheme.

\subsection{Importance-aware Token-level Alignment}

To address the gradient dilution issue, CORD introduces a multi-dimensional weighting scheme that prioritizes semantically critical and early-stage tokens.

\paragraph{Top-$K$ KL-based Importance Weighting.}
We first select the $K$ tokens with the largest divergence values, where $K=20$ in all experiments. Let $\mathcal{I}^*_K$ denote the index set of the top-$K$ tokens ranked by $D_t$. We define a KL-based importance weight 
\begin{equation} 
w_t^{\mathrm{KL}} = \begin{cases} \alpha, & t \in \mathcal{I}^*_K, \\ 
1, & \text{otherwise}, \end{cases} 
\label{eq:topk_weight} 
\end{equation} 
where $\alpha > 1$ is a hyperparameter that controls the strength of emphasis on high-divergence tokens. This hard selection prevents low-divergence tokens from dominating the optimization and ensures that gradient updates focus on tokens that exhibit significant cross-modal semantic mismatch.

\paragraph{Sequential Decay Weighting.}
We further observe that early decoding errors are more detrimental than later ones: once an incorrect semantic decision is made at an early step, subsequent tokens are unlikely to fully correct the reasoning trajectory. To emphasize early alignment, we introduce a position-dependent decay weight. Specifically, we assign a higher weight to earlier tokens and linearly decay it over time: 
\begin{equation} w_t^{\mathrm{pos}} = \beta - (\beta - 1)\frac{t-1}{T-1}, \qquad t = 1,\ldots,T, 
\label{eq:position_weight} 
\end{equation} 
where $\beta > 1$ is a hyperparameter controlling the relative importance of early decoding steps. This formulation ensures that $w_1^{\mathrm{pos}} = \beta$ and $w_T^{\mathrm{pos}} = 1$.

\paragraph{Final Token-level Alignment Objective.}
The final token weight is defined as the product of the KL-based importance weight
and the positional decay weight:
\begin{equation}
w_t = w_t^{\mathrm{KL}} \cdot w_t^{\mathrm{pos}}
\label{eq:final_weight}
\end{equation}

The token-level alignment loss is then given by
\begin{equation}
\mathcal{L}_\mathrm{tok}
=
\mathbb{E}_{y \sim p_\theta(\cdot \mid x^a)}
\left[
\sum_{t=1}^{T}
w_t , D_t
\right]
\label{eq:token_loss}
\end{equation}

This objective amplifies supervision on a small set of semantically critical tokens
while prioritizing early decoding stages, effectively mitigating gradient dilution
and improving cross-modal reasoning alignment.

\subsection{Sequence-level Alignment via Reward-guided Optimization}

While token-level alignment corrects local semantic deviations, it does not explicitly
constrain global reasoning behavior.
In practice, locally aligned token distributions may still lead to globally inconsistent
or incorrect final answers.
To address this limitation, CORD introduces a sequence-level alignment objective that
provides global supervision over complete audio-conditioned reasoning trajectories.

\paragraph{Judge-based Global Alignment Reward.}
Given an audio input $x^{a}$ and its semantically equivalent text input $x^{t}$,
we sample an audio-conditioned output sequence
$y \sim p_{\theta}(\cdot \mid x^{a})$ and generate a text-conditioned reference
$\hat{y} \sim p_{\theta}(\cdot \mid x^{t})$.
We employ a judge model $J(\cdot,\cdot)$ to directly evaluate whether the two sequences
are semantically aligned at the answer level.
The judge produces a binary reward:
\begin{equation}
r_{\text{seq}}(y; x^{a}, x^{t})
=
J(y, \hat{y})
\in \{0,1\},
\label{eq:judge_reward}
\end{equation}
where $r_{\text{seq}}=1$ indicates that the audio-conditioned answer is judged to be
semantically consistent with the text-conditioned answer, and $r_{\text{seq}}=0$
otherwise.
This reward captures global reasoning alignment beyond local token-wise similarity.

\newcommand{\deltasm}[1]{\textnormal{\scriptsize(#1)}}

\newcolumntype{C}[1]{>{\centering\arraybackslash}p{#1}}

\begin{table*}[t]
\centering
\small
\setlength{\tabcolsep}{8pt} 
\renewcommand{\arraystretch}{1.3}
\begin{tabular}{l *{3}{c C{4.5em}} c} 
\toprule
\multirow{2}{*}{\textbf{Method}}
& \multicolumn{2}{c}{\textbf{MMSU}}
& \multicolumn{2}{c}{\textbf{OBQA}}
& \multicolumn{2}{c}{\textbf{GSM8K}}
& \multirow{2}{*}{\textbf{AVG.}} \\
\cmidrule(lr){2-3}\cmidrule(lr){4-5}\cmidrule(lr){6-7}
& {Acc.} & \multicolumn{1}{c}{$\Delta_{\text{base}}$} 
& {Acc.} & \multicolumn{1}{c}{$\Delta_{\text{base}}$}
& {Acc.} & \multicolumn{1}{c}{$\Delta_{\text{base}}$} \\
\midrule
\textbf{Qwen2-Audio-7B-Instruct}
& 36.04 & 8.42
& 51.20 & 17.59
& 20.73 & 19.74
& 15.25 \\
\ \ + SFT
& 36.47 & 7.99 \down{0.43}
& 49.49 & 19.30 \up{1.71} 
& 33.68 & 6.79 \down{12.95}
&  11.36 \\
\ \ + Forward KL
& 36.77 & 7.69 \down{0.73}
& 48.21 & 20.58 \up{2.99} 
& 36.05 & 4.42 \down{15.32}
& 10.90 \\
\rowcolor{pyellow}
\ \ + CORD \textit{(ours)}
& \textbf{38.06} & \textbf{6.40} \textbf{\down{2.02}}
& \textbf{52.77} & \textbf{16.02} \textbf{\down{1.57}}
& \textbf{36.20} & \textbf{4.27} \textbf{\down{15.47}}
& \textbf{8.90} \\
\midrule
\textbf{Step-Audio2 mini}
& 52.31 & 8.16
& 72.30 & 7.70
& 43.75 & 16.72
& 10.86 \\
\ \ + SFT
& 54.68 & 5.79 \down{2.37}
& 74.72 & 5.28 \down{2.42}
& 27.89 & 32.58 \up{15.86}
& 14.55 \\
\ \ + Forward KL
& 53.12 & 7.35 \down{0.81}
& 72.08 & 7.92 \up{0.22} 
& 46.57 & 13.90 \down{2.82}
& 9.72 \\
\rowcolor{pyellow}
\ \ + CORD \textit{(ours)}
& \textbf{57.63} & \textbf{2.84} \textbf{\down{5.32}}
& \textbf{77.74} & \textbf{2.26} \textbf{\down{5.44}}
& \textbf{47.56} & \textbf{12.91} \textbf{\down{3.81}}
& \textbf{6.00} \\
\bottomrule
\end{tabular}
\caption{Comparison of audio-conditioned reasoning performance across different backbones. \textbf{Base Models} (first row of each block) refer to the original instruction-tuned Qwen2-Audio-7B and Step-Audio2-mini. $\Delta_{\text{base}}$ represents the modality gap for each method, defined as $\text{Acc}_{\text{text}}^{\text{Base}} - \text{Acc}_{\text{audio}}^{\text{Method}}$, where $\text{Acc}_{\text{text}}^{\text{Base}}$ is the fixed text-conditioned accuracy of the Base Model. For Qwen2-Audio, these text baselines are 44.46, 68.79, and 40.47; for Step-Audio2, they are 60.47, 80.00, and 60.47. \textbf{AVG.} is the mean of the modality gaps. Arrows indicate the reduction ({\color{darkgreen}$\downarrow$}) or increase ({\color{red}$\uparrow$}) in the gap compared to the Base Model’s initial $\Delta_{\text{base}}$.}

\label{tab:text_audio_gap}
\end{table*}

\paragraph{GRPO Optimization.}
To optimize the model under this sequence-level reward, we adopt
Group Relative Policy Optimization.
For each audio input $x^{a}$, we sample a group of $N$ on-policy trajectories
$\{y^{(i)}\}_{i=1}^{N}$ from the current policy $p_{\theta}(\cdot \mid x^{a})$.
Each trajectory is evaluated by the judge model, yielding rewards
$\{r_{\text{seq}}^{(i)}\}_{i=1}^{N}$.

GRPO computes a relative advantage by comparing each trajectory’s reward to the
group average:
\begin{equation}
A^{(i)}
=
r_{\text{seq}}^{(i)}
-
\frac{1}{N}
\sum_{j=1}^{N}
r_{\text{seq}}^{(j)} .
\label{eq:grpo_advantage}
\end{equation}

Following the approach in DAPO \cite{yu2025dapoopensourcellmreinforcement}, we omit the explicit KL divergence penalty. Consequently, the sequence-level optimization objective is defined as:
\begin{equation}
\mathcal{L}_{\text{seq}} = - \mathbb{E}_{\{y^{(i)}\}_{i=1}^N} \left[ \frac{1}{N} \sum_{i=1}^{N} A^{(i)} \log p_{\theta}(y^{(i)} \mid x^{a}) \right]
\label{eq:grpo_loss}
\end{equation}

By optimizing this objective, the model increases the likelihood of audio-conditioned
trajectories that achieve higher global alignment rewards relative to other on-policy
rollouts.
Crucially, GRPO operates entirely on-policy, ensuring that global supervision is applied
to the exact inference states encountered during audio-conditioned reasoning.

Together with token-level alignment, this judge-guided GRPO objective explicitly
constrains global reasoning outcomes and mitigates failure cases where locally aligned
tokens still result in semantically inconsistent answers.

\subsection{Overall Objective}
CORD jointly optimizes local and global alignment objectives:
\begin{equation}
\mathcal{L}_{\text{CORD}} = \mathcal{L}_{\text{tok}} + \mathcal{L}_{\text{seq}},
\end{equation}
where $\mathcal{L}_{\text{seq}}$ denotes the GRPO-based sequence-level loss.

\section{Experiments}

\subsection{Experimental Setting}

\paragraph{Baselines.}

In our experiments, we adopt \texttt{Qwen2-Audio-7B-Instruct} \citep{chu2024qwen2} and \texttt{Step-Audio-2-Mini} \citep{wu2025step} as base models. Following the previous distillation method~\citep{wang2024blspbootstrappinglanguagespeechpretraining,cuervo2025closing}, We implement two distillation objectives using teacher rollouts:

\begin{itemize}
    \item \textbf{Supervised Fine-Tuning (SFT):} Given a teacher rollout $y \sim p_{\theta}(\cdot \mid x^t)$, the student is optimized to maximize the likelihood of the audio-conditioned trajectory:
    \begin{equation}
    \mathcal{L}_{\text{SFT}} = -\mathbb{E}_{y \sim p_{\theta}^t} \left[ \sum_{t=1}^{|y|} \log p_{\theta, t}^a \right]
\end{equation}

    \item \textbf{Forward KL Divergence:} We minimize the distribution discrepancy between modalities over teacher rollouts:
    \begin{equation}
        \mathcal{L}_{\text{FKL}}
        =
        \mathrm{KL} \left(
        p_{\theta}(\cdot \mid y_{<t}, x^t)
        \,\middle\|\,
        p_{\theta}(\cdot \mid y_{<t}, x^{a})
        \right)
    \end{equation}
\end{itemize}

\paragraph{Dataset.}
We curate 80,000 instances from \textbf{NuminaMath}~\citep{numina_math_datasets} to construct our training data.
By employing the Kokoro~\citep{kokoro2024} model to synthesize audio for each text instruction, we generate semantically equivalent input pairs across the \emph{text} and \emph{audio} modalities.

We intentionally train on a single dataset to keep the training distribution controlled, so that improvements can be attributed to the proposed cross-modal alignment objectives rather than to broader data diversity.
We choose NuminaMath primarily because it offers a large collection of math problems that elicit step-by-step reasoning traces, and its text prompts can be directly converted into audio with minimal ambiguity, making it convenient for constructing paired audio--text inputs at scale.

Despite being trained on a math-only corpus, CORD improves performance on non-mathematical benchmarks such as MMSU and OpenBookQA (Section~\ref{sec:main_results}), suggesting that it learns modality-bridging alignment beyond dataset-specific patterns.
We expect that incorporating more diverse instruction data (e.g., general QA, knowledge-intensive reasoning, and non-speech acoustic tasks) would further strengthen cross-modal generalization, which we leave for future work.

\paragraph{Judge Model.} Our judge model for GRPO was developed by distilling the evaluation outputs of proprietary frontier models on millions of text-based instruction-following samples. This judge model exhibits exceptional performance, with self-evaluation accuracy consistently exceeding 99\%.

\paragraph{Evaluation Benchmarks.}
To evaluate the model's knowledge-based question answering capabilities, we utilize MMSU and OpenBookQA (OBQA) from VoiceBench \cite{voicebench}. GSM8K~\citep{cobbe2021gsm8k} is employed to assess the model's mathematical reasoning performance. Furthermore, we adopt MMAU \cite{mmau} to evaluate acoustic and paralinguistic reasoning; this benchmark is further categorized into three distinct subsets: \textit{Music}, \textit{Speech}, and \textit{Sound}.

\paragraph{Experimental Details.}
For CORD and Forward-KL distillation baselines, we generate a single on-policy rollout per prompt at each update step.
In contrast, for GRPO, we sample a group of four rollouts per prompt (group size $N{=}4$).
The maximum generated sequence length is set to 200 tokens.
All models are optimized using the AdamW optimizer with a learning rate of $3\times10^{-5}$ and share the same training schedule and decoding configurations unless otherwise specified.
For rollout sampling, we use a temperature of 1.0 for CORD and Forward-KL, while a higher temperature of 1.5 is adopted for GRPO to encourage trajectory diversity. For the token-level alignment objective, we fix the KL-based importance
scaling factor and the positional decay factor to $\alpha = 2$ and
$\beta = 2$, respectively, in all experiments.

\subsection{Main Results} \label{sec:main_results}

Tables~\ref{tab:text_audio_gap} and~\ref{tab:mmau_modality} report the performance of CORD across multiple benchmarks using two representative LALMs, Qwen2-Audio-7B-Instruct and Step-Audio2-mini.
The results show that CORD substantially \textbf{reduces the performance gap} between audio and text modalities, while effectively \textbf{preserving auxiliary audio capabilities} beyond speech.


\paragraph{Cross-modal gap reduction.}
As shown in Table~\ref{tab:text_audio_gap}, CORD achieves consistent and substantial reductions in the audio–text performance gap across both backbone models.
Across all benchmarks, CORD consistently outperforms SFT and Forward-KL, demonstrating the effectiveness of on-policy, trajectory-level alignment. 
Specifically, on Qwen2-Audio-7B-Instruct, CORD reduces the average audio–text gap by \textbf{41.6\%} relative to the base model, whereas Forward-KL yields only a 28.5\% reduction.
On Step-Audio2-mini, CORD achieves a \textbf{44.8\%} gap reduction, while Forward-KL leads to only a 10.5\% reduction.

In addition, by comparing results between different models, we observe a clear trend: Step-Audio2-mini benefits more from CORD ($44.8\% \rightarrow 41.6\%$), exhibiting larger gap reductions and stronger alignment. 
This suggests that a stronger text-conditioned teacher leads to more effective cross-modal alignment. 
This observation indicates that CORD naturally scales with the quality of the base LALM and can more effectively exploit improvements in text reasoning to enhance audio-conditioned performance.

\paragraph{Emergent Cross-domain Generalization.}
Although CORD is trained exclusively on a math-focused dataset, both models exhibit more pronounced improvements on general-domain benchmarks such as MMSU and OBQA than on the math-intensive GSM8K benchmark. This observation indicates that CORD does not merely acquire domain-specific knowledge, but instead learns a transferable meta-capability of cross-modal alignment.

\begin{table}[t]
\centering
\small
\renewcommand{\arraystretch}{1.3}
\begin{tabular}{lcccc}
\toprule
\textbf{Method} & \textbf{Music} & \textbf{Sound} & \textbf{Speech} & \textbf{Avg.}\\
\midrule
\textbf{Base Model} & 58.98 & 64.74 & 58.73 & 60.81\\
\   + SFT               & 56.29 & 64.44 & 51.51 & 57.39 \\
\   + Forward KL        & 55.99 & 61.70 & 53.01 & 56.90\\
\rowcolor{pyellow} 
\  + CORD \textit{(ours)  }     & 60.18 & 64.44 & 55.42 & 60.01\\
\bottomrule
\end{tabular}
\caption{Fine-grained performance on the MMAU benchmark across music, sound, and speech categories based on Qwen2-Audio-7B-Instruct.}
\label{tab:mmau_modality}
\end{table}

\paragraph{Preserving Auxiliary Audio Capabilities.}
Table~\ref{tab:mmau_modality} shows that CORD effectively preserves general audio understanding capabilities while aligning for complex reasoning, whereas baseline methods suffer from significant performance degradation. Specifically, the Forward-KL baseline exhibits a noticeable performance tax, with scores dropping by 2.99 points in \emph{music} and 3.04 points in \emph{sound}, suggesting that conventional distillation may inadvertently lead to the catastrophic forgetting of non-speech acoustic patterns. In contrast, CORD demonstrates remarkable robustness: it not only maintains near-parity with the base model in the \emph{sound} category but even yields a slight improvement in \emph{music}. These results indicate that on-policy alignment effectively mitigates collateral damage to auxiliary audio modalities, ensuring a more stable and balanced cross-modal alignment that retains pre-trained general audio intelligence.

\subsection{Ablation Studies}
\label{sec:ablation}

We conduct ablation studies to analyze the contribution of each component in CORD and to investigate the sensitivity of key hyperparameters.
All ablations are performed on Qwen2-Audio-7B-Instruct under the same training setup as the main experiments.

\subsubsection{Component-wise Ablation}

In Table~\ref{tab:ablation_components}, we conduct an incremental ablation study to evaluate the efficacy of each module within the CORD framework. 

We observe that while initial reinforcement learning via GRPO yields performance gains at 500 steps, it suffers from severe model collapse as training extends to 1000 steps. This degradation is most pronounced on GSM8K (35.59 $\rightarrow$ 19.89), where performance falls even below the base model baseline. 
Notably, our OPD module acts as a powerful regularizer that eliminates this instability, ensuring sustained optimization for up to 3000 steps without quality loss (35.59 $\rightarrow$ 36.12 in GSM8K). 
Finally, by augmenting OPD with cross-modal token-level weighting, the complete CORD framework achieves the best overall performance, with an average improvement of 6.35 over the base model (i). 
This validates the synergy between stable sequence-level reinforcement learning, on-policy anchoring, and fine-grained alignment.

\begin{table}[t]
\centering

\small 
\renewcommand{\arraystretch}{1.2}

\setlength{\tabcolsep}{5pt}

    \begin{tabular}{l c ccc}
    \toprule
    \textbf{Method} & \textbf{Step} & \textbf{MMSU} & \textbf{OBQA} & \textbf{GSM8K} \\
    \midrule
    Base Model (i) & -- & 36.04 & 51.20 & 20.73 \\
    \midrule
    \rowcolor{gray!10} \multicolumn{5}{l}{\textit{Stability Analysis (GRPO only)}} \\
    \ \ + GRPO (ii) & 500 & 36.92 & 50.54 & 35.59 \\
    \ \ + GRPO (ii) & 1000 & \textcolor{red}{27.87} & \textcolor{red}{36.48} & \textcolor{red}{19.89} \\
    \midrule
    \rowcolor{gray!10} \multicolumn{5}{l}{\textit{Incremental Ablation (Cumulative)}} \\
    \ \ + GRPO + OPD (iii )& 3000 & 37.41 & 51.20 & 36.12 \\
    \rowcolor{pyellow}
    \ \ + GRPO + OPD + Weight (iv) & 3000 & \textbf{38.06} & \textbf{52.77} & \textbf{36.20} \\
    \bottomrule
    \end{tabular}

\caption{Ablation study of individual components and training stability. We evaluate the incremental contributions of: (i) the base model, (ii) sequence-level \textbf{GRPO}, (iii) On-Policy Distillation (\textbf{OPD}), and (iv) the Full CORD framework, which integrates all previous modules with \textbf{token-level importance weighting} for fine-grained alignment. \textcolor{red}{Red values} denote model collapse where performance falls below the base baseline.}
\label{tab:ablation_components}
\end{table}

\begin{figure}[t]
\centering
\small
\begin{tikzpicture}
\begin{axis}[
    width=0.95\linewidth,
    height=5.2cm,
    xlabel={Weighting Intensity ($\alpha = \beta$)},
    ylabel={Relative $\Delta$ Score},
    xtick={1.0, 1.5, 2.0, 2.5},
    xmin=0.9, xmax=2.6,
    ymin=-0.1, ymax=2.1,
    ytick={0, 0.5, 1.0, 1.5, 2.0},
    grid=both,
    grid style={dashed, gray!20},
    legend style={
        at={(0.5, 1.02)},
        anchor=south,
        legend columns=3,
        draw=none,
        fill=none,
        font=\footnotesize,
        /tikz/every even column/.append style={column sep=10pt}
    },
    tick label style={font=\footnotesize},
    label style={font=\footnotesize},
    axis line style={gray!60},
    tick align=outside
]

\addplot[
    very thick,
    color=blue!60!cyan,
    mark=o,
    mark size=2pt,
    mark options={fill=white, solid}
] coordinates {
    (1.0, 0.00) (1.5, 0.13) (2.0, 0.91) (2.5, 0.64)
};
\addlegendentry{MMSU}

\addplot[
    very thick,
    color=orange!90!black,
    mark=square*,
    mark size=1.8pt,
    mark options={solid}
] coordinates {
    (1.0, 0.00) (1.5, 1.10) (2.0, 1.76) (2.5, 1.78)
};
\addlegendentry{OBQA}

\addplot[
    very thick,
    color=green!50!black,
    mark=triangle,
    mark size=2.2pt,
    mark options={fill=white, solid}
] coordinates {
    (1.0, 0.00) (1.5, 0.07) (2.0, 0.30) (2.5, 0.28)
};
\addlegendentry{GSM8K}

\draw[dashed, gray!80, thick] (axis cs:2.0, -0.1) -- (axis cs:2.0, 2.1);
\node[anchor=south west, font=\tiny, gray!80] at (axis cs:2.0, 0.05) {Optimal};

\end{axis}
\end{tikzpicture}
\caption{Sensitivity analysis of the weighting intensity $\alpha$ and $\beta$. To reduce hyperparameter complexity, we set $\alpha = \beta$. The scores represent relative improvements over the baseline ($\alpha=\beta=1.0$). A value of $2.0$ yields the most consistent gains across all tasks.}
\label{fig:alpha_beta_sensitivity}
\end{figure}
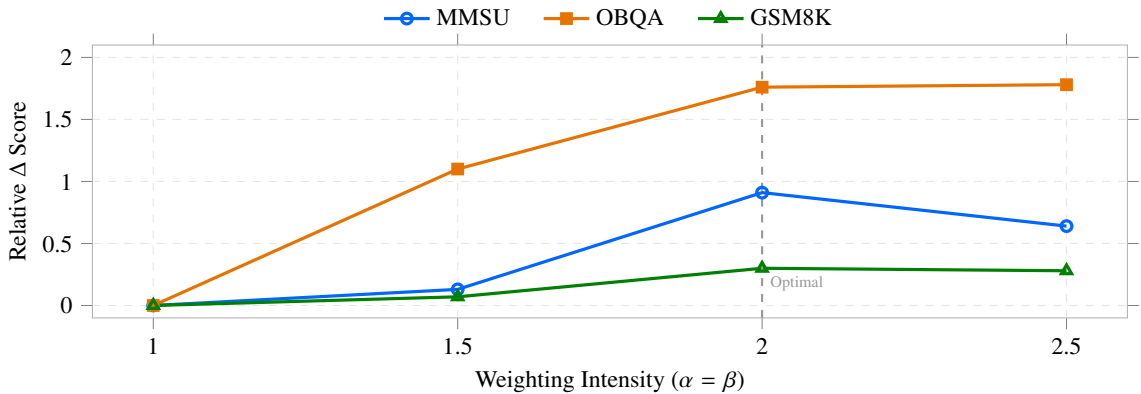

\subsubsection{Sensitivity to Weighting Intensity}
To efficiently navigate the hyperparameter space while balancing the contributions of token-level importance ($\alpha$ in Equation.~\ref{eq:topk_weight} ) and positional weighting ($\beta$ in Equation.~\ref{eq:position_weight} ), we couple these parameters by setting $\alpha = \beta$. As illustrated in Figure~\ref{fig:alpha_beta_sensitivity}, the performance across all benchmarks exhibits a characteristic bell-shaped trend, achieving an optimal trade-off at $\alpha = \beta = 2.0$. This configuration sufficiently accentuates pivotal reasoning states without over-concentrating gradients on a sparse subset of tokens, which could otherwise compromise optimization stability. 
In contrast, smaller values (e.g., 1.0) revert toward uniform KL optimization, while excessively large values (e.g., 2.5) lead to marginal performance drops due to the over-suppression of long-tail semantic information. 
Consequently, we adopt $\alpha = \beta = 2.0$ as the default setting for all primary experiments.

\section{Conclusion}

We introduce \textbf{CORD}, a weighted on-policy cross-modal self-distillation framework designed to bridge the audio--text gap in LALMs. By aligning audio-conditioned reasoning with text baselines at both token and sequence levels, CORD employs an importance-aware KL objective and a judge-guided GRPO objective to ensure local semantic accuracy and global trajectory consistency. Extensive benchmarks demonstrate that CORD significantly narrows the modality gap, highlighting on-policy trajectory alignment as a robust paradigm for cross-modal alignment.

\clearpage
\bibliography{ref}
\bibliographystyle{colm2024_conference}

\clearpage
\appendix

\end{document}